\documentstyle[11pt,appb,epsfig,float,tables]{article} 

\preprint{} 
\date{0} 

\def\beq{\begin{equation}}
\def\eeq{\end{equation}}
\def\bea{\begin{eqnarray}}
\def\eea{\end{eqnarray}}
\def\MSbar {\hbox{$\overline{\hbox{MS}}\,$}}
\newcommand{\eqref}[1]{(\ref{#1})}   

\begin{document}
\title{
\begin{flushright}
$\phantom{a}$\vskip-7em
{\normalsize TAUP 2493-97}
\end{flushright}
\vskip3em
PRECISE ESTIMATES OF HIGH ORDERS IN QCD\thanks{Invited talk at 
the Cracow Epiphany Conference on
Spin Effects in Particle Physics, January 9-11, 1998, Cracow, Poland;}
}
\author{Marek KARLINER
\address{
School of Physics and Astronomy
\\ Raymond and Beverly Sackler Faculty of Exact Sciences
\\ Tel-Aviv University, 69978 Tel-Aviv, Israel
\\ e-mail: marek@proton.tau.ac.il
\\}
}
\maketitle
\begin{abstract}
I review the
recent work on obtaining precise estimates of higher-order corrections 
in QCD and field theory.
\end{abstract}
  
\section{Introduction}
The precision of the experimental data on electroweak
interactions and QCD
is now very high and it is expected to become significantly
higher within the next few years.
This has triggered a substantial refinement in
the corresponding theoretical calculations. Yet, already now
for certain experimental quantities
the theoretical uncertainty is one of the
major open questions in the interpretation of the data and in
the search for signals of physics beyond the Standard Model.
A striking example is the need for a precise
determination of the gauge couplings at the weak scale,
which is the prerequisite for investigation of possible
unification of couplings at some GUT scale.
 
One of the reasons for this current state of affairs
in the relation between the theory and experiment
is that
computation of high orders in perturbation theory
for quantum field theories, and especially non-abelian
gauge theories
in $3{+}1$ dimensions is extremely hard.
State-of-the-art
calculations available today for this kind of theories
have reached, after a very large effort, the 3-rd and the 4-th order 
in $\alpha_s$, 
for observables and for the $\beta$-function, respectively
\cite{high_orders,beta4loop,massAnDim}.
Without a major breakthrough in
the relevant techniques it is unlikely that exact results for  the next
order will become available in the foreseeable future.
Moreover, even if explicit expressions for
very high order terms do become available, we still
have to deal with the fact that
the perturbative series of interest
are asymptotic, with zero radius of convergence  and usually are not even 
Borel summable. In this talk I will review an approach which has been
recently suggested to deal with some of these problems.


\section{Perturbation Theory: Diseases and a Promising Therapy}
As mentioned in the Introduction,
    the perturbation series in QCD is expected to be asymptotic with
rapidly growing coefficients:
\beq
S(x) = \sum^\infty_{n=0} c_n x^n~, \quad x \equiv \frac{\alpha_s}{\pi}~,
\qquad c_n
\simeq n!K^n {n}^\gamma
\label{GenericSeries}
\eeq
for some coefficients $K, \gamma$ \cite{renmvz,RenormRev}.
Anyone who wants to make use of QCD perturbation theory to carry out
precision analysis of observables has to face several practical problems:

\begin{itemize}
\item only few first orders in (\ref{GenericSeries}) are known for any
observable ($n\le3$)
\item the series has zero radius of convergence 
\item the series is usually not {\em Borel summable}.
Borel summation is a trick that sometimes works for summing 
series with factorial divergence.
Consider the series for $S(x)$ in eq.~(\ref{GenericSeries}).
We can define a new function, ${\cal S}(y)$, whose series is obtained
from (\ref{GenericSeries}) by dividing the $n$-th term by $n!$,
\beq
{\cal S}(y) = \sum^\infty_{n=0} \left({c_n\over n!}\right) y^n
\label{BorelSeries}
\eeq
If the new series is convergent, the original function $S(x)$ can be
obtained by the so-called inverse Borel transform,
\beq
S(x) = {1\over x}\,\int_0^\infty e^{{-}y/x} {\cal S}(y) dy
\label{invBorelTransform}
\eeq
provided ${\cal S}(y)$ has no singularities along the integration path.
\vskip0.5truecm
\parbox{6.5truecm}{
\makebox[6.5truecm][l]{\epsfig{file=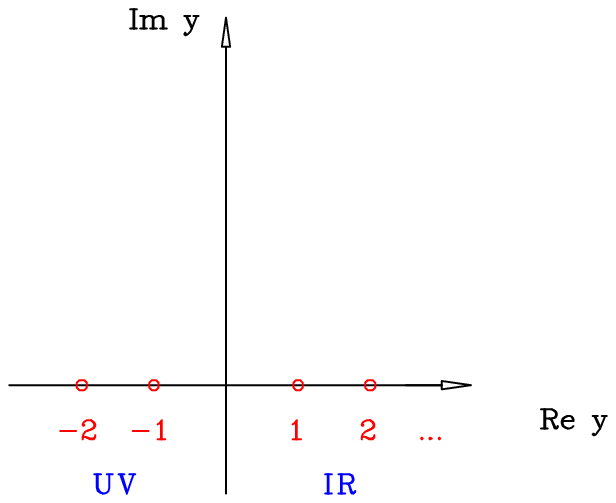,
width=6.0truecm,angle=0}}
}
\parbox{5truecm}{
Unfortunately, in QCD it is known that for a generic observable
${\cal S}(y)$ has poles on both the positive {\em and} negative $y$ axis
in the complex $y$ plane. These are usually referred to as {\em infrared} and 
{\em ultraviolet renormalons}, respectively. 
The figure to the left shows a
schematic description of poles in the Borel transform of a generic
series for a QCD observable.
}

The presence of singularities along the integration path makes the integral 
(\ref{invBorelTransform}) ill-defined. One can try to define it by going
around the poles, but this introduces an ambiguity proportional to the
pole residue, since different deformations of the integration path
will give different results.

\item
renormalization scale dependence: 
finite-order perturbative predictions depend on the arbitrary 
renormalization scale $\mu$ through the 
coupling,
$\alpha_s=\alpha_s(\mu)$.
This renormalization scale is most pronounced at leading order in
perturbation theory and
decreases with the inclusion of higher order terms.

\item
renormalization scheme dependence: in principle, the theory can be
renormalized in any valid renormalization scheme, yielding the same 
predictions for any physical observable.
 In practice, when we work
with a finite number of perturbative terms, the results depend on the
renormalization scheme.
\end{itemize}

\noindent
There is no ``miracle cure" which would solve these problems completely.
However, we should and can minimize their effect. Thus the practical
issue is how to get the best possible precision, given a fixed number of
terms in the perturbative expansion. In the following, I will discuss one
method which has already shown considerable progress towards this goal.
The method is based on the so-called Pad\'e Approximants (PA-s)
\cite{PadeInQFT}-\cite{EJJKS}.


\section{Pad\'e to the Rescue}

    Pad\'e Approximants \cite{Baker,BenderOrszag}
are rational functions chosen to have a Taylor expansion equal the
perturbative series to the order calculated. Given a series
\beq
S(x) = c_0 + c_1 x + c_2 x^2 +c_3 x^3 + \cdots 
+c_N x^N +\,{\cal O}(x^{N+1})
\label{StoN}
\eeq
one can always find a rational function 
\beq
[L/M] \equiv \frac{a_0 + a_1x + ... +a_Lx^L}{1 + b_1x + ... + b_Mx^M}~:
\qquad L+M = N
\label{PadeDef}
\eeq
such that $[L/M]$ has the Taylor series
\beq
[L/M] =  c_0 + c_1 x + c_2 x^2 +c_3 x^3 + \cdots
+c_N x^N + \,\tilde c_{N+1} x^{N+1} + \cdots
\label{LMseries}
\eeq
The rational function in (\ref{PadeDef}) is called
the $[L/M]$ Pad\'e Approximant.

It is important to keep in mind that at a given finite order in $x$
the $[L/M]$ Pad\'e in (\ref{PadeDef}) is 
formally as valid representation of $S(x)$
as the original perturbation expansion. 
Moreover, in practice the PA-s turn out to posses many important and 
useful properties which
are absent in the straightforward perturbation theory.

Thus, even though PA is constructed to reproduce the series
\eqref{StoN} only up to order $N$, it turns out that
under rather mild conditions the next term in the Taylor 
expansion of the PA in eq.~\eqref{LMseries}, 
$\tilde c_{N+1}$,
provides a good estimate,
$c^{est}_{N+1}=\tilde c_{N+1}$.
We call it the {\em Pade Approximant Prediction} (PAP),
of the next coefficient $c_{N+1}$ in the series
\eqref{StoN}:
\beq
\left\vert \frac{c^{est}_{N+1}}{c_{N+1}} -1 \right\vert \ll 1
\label{estI}
\eeq
and for sufficiently large $N$ the relative error decays 
exponentially fast,
\beq
\left\vert \frac{c^{est}_{N+1}}{c_{N+1}} -1 \right \vert \sim 
e^{-\sigma N}\,,\qquad \sigma=\hbox{const}
\label{estII}
\eeq
Let us consider some simple examples, starting with 
the trivial case of a single-pole geometric series
\beq
\frac{A}{1 - B x} = \sum_{n=0}^\infty c_n x^n
\label{singlepole}
\eeq
It is easy to convince oneself that in this case the $[L/M]$ Pad\'e
is {\em exact} for $L\ge 0$, $M\ge 1$. For example, if we attempt to 
construct a [10/10] Pad\'e of \eqref{singlepole}, we will find that 
the {\em a priori} 10-th degree polynomials in numerator and denominator
reduce to a degenerate case of a constant and 1-st degree polynomial, 
respectively,
\beq
[10/10] \equiv \frac{P_{10}(x)}{Q_{10}(x)} = \frac{A}{1 - B x}
\label{tenten}
\eeq
Once this is clear, the extension to a sum of finite number of poles in
obvious,
\beq
\sum_{i=1}^K \frac{A_i}{1 - B_i x} = \sum_{n=0}^\infty c_n x^n
\qquad \Longrightarrow \qquad [L/M] \hbox{\ exact for\ } 
\,L \ge K-1,\ 
M \ge  K
\label{Npoles}
\eeq
One can also show that for an infinite number of isolated poles, i.e. when
$f(x) = \displaystyle\sum_0^\infty c_n x^n $ is a meromorphic function,
the sequence of $[L/L+k]$ for $k$ fixed converges to $f(x)$ as
$L\rightarrow \infty$,
\beq
[L/L+k] \,\,
\mathop{\hbox to 2.7em{\rightarrowfill}}_{L\rightarrow \infty}
\,\, f(x)\,; \qquad\,
k = 0, \pm 1, \pm 2,\dots 
\eeq
A somewhat less intuitive, but very important result is that 
in certain cases the Pad\'e sequence $[L/L+k]$ converges 
exponentially fast in $L$ to the 
correct function even for a factorially divergent
asymptotic series with zero radius of
convergence. A classical example \cite{BenderOrszag} is the function
\beq
g(x) = \int_0^\infty \frac{e^{-t}}{1 + x t} \, d t = 
\sum_0^\infty (-x)^n\,n!
\label{factorialseries}
\eeq
Here again it turns out that 
$[L/L+k] \, \longrightarrow \, g(x)$ as $L \rightarrow \infty$.

The crucial property of the series in \eqref{factorialseries}
which makes this possible
 is that 
it has alternating signs. It is easy to show that this implies that 
all the poles of the Borel transform of \eqref{factorialseries}
are on the negative real axis, and hence that the series is Borel summable.
More generally, when the series is Borel summable, Pad\'e will converge
to the correct result.

\begin{figure}[H]
\begin{center}
\mbox{\epsfig{file=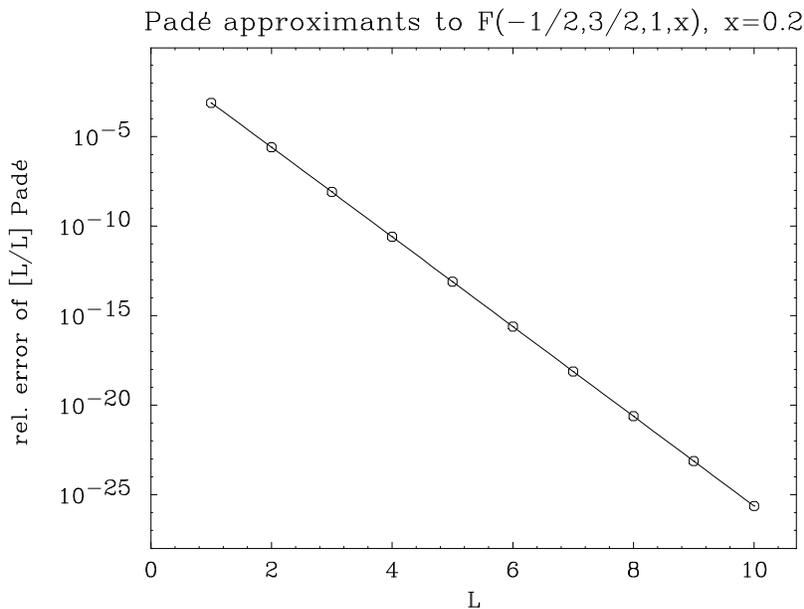,width=8.0truecm,angle=90}}
\end{center}
\caption{\small 
Exponentially fast
convergence of Pad\'e despite the presence of a cut:
relative error of diagonal [L/L] Pad\'e for the 
hypergeometric function $F(-1/2,3/2,1,x)$ which has a cut
for $x \ge 1$. The Pad\'e is evaluated to the left of the cut,
at $x=0.2$.
\label{hyperfig}
}   
\end{figure}

It is interesting to note that the exponentially fast convergence 
of PA-s is not limited to meromorphic functions. As a simple example, 
consider the hypergeometric function $F({-}{1\over2},{3\over2},1,x)$
\cite{Abramowitz},
which has a cut for $x\ge1$. Fig.~\ref{hyperfig}
shows that despite the cut, the diagonal PA-s evaluated at 
$x=0.2$ converge exponentially fast.
\bigskip

\section{Applications to Quantum Field Theory}
\begin{figure}[H]
\begin{center}
\mbox{\epsfig{file=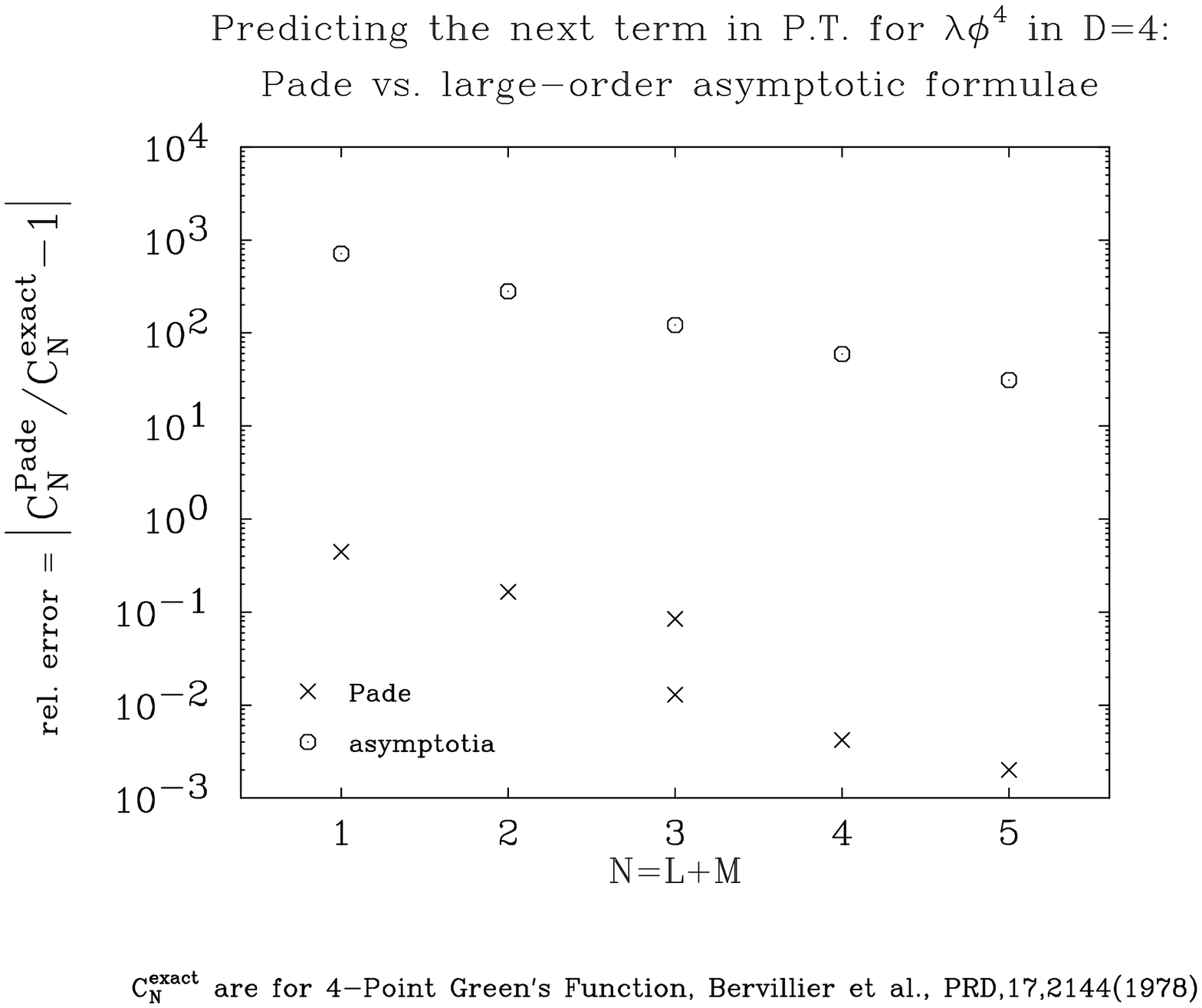,width=9.0truecm,angle=90}}
\end{center}
\caption{\small
Pad\'e predictions (crosses)
for high-order terms in the perturbative expansion
for 4-point Green's function in a scalar field theory in $D=4$.
For comparison also included are 
predictions based on asymptotic large-order behavior (circles).
\label{phi4fig}
}
\end{figure}

While such mathematical examples are instructive, in order to gain
confidence in the method, we need to see how it fares on high-order series
taken from quantum field theory. As the first test case, we consider 
the scalar field theory with Gaussian propagators. 
High-order perturbation expansions of Green's functions in this theory
have been computed in Ref.~\cite{Bervillier}.
Fig.~\ref{phi4fig} demonstrates the convergence of PAP for the relevant
coefficients for the 4-point Green's function in $D=4$ \cite{phi4PAP}.
The relative error is
$\sim 10^{-3}$ at 5-th order. For comparison also shown are relative errors
of estimates based on asymptotic behavior of large orders in perturbation
theory, as given in \cite{phi4PAP}.
Clearly, at 5-th order Pad\'e does better by about 4 orders 
of magnitude.
\bigskip

If information is available about the asymptotic behavior of
$c_n$, it is possible to obtain an explicit expression for the
the error formula on the
r.h.s. of \eqref{estII}.
  For
example, we have demonstrated that if
\beq
\epsilon_n \equiv \frac{c_{n}\, c_{n+2}}{c^2_{n+1}} - 1
\,\simeq \,{1\over n},
\label{epsDef}
\eeq
as is the case for any series dominated by a finite number of
renormalon singularities, then $\delta_{[L/M]}$ defined by
\beq
\delta_{[L/M]} \equiv \frac{c^{est.}_{L+M+1} -
c_{L+M+1}}{c_{L+M+1}}
\label{Efour}
\eeq
has the following asymptotic behaviour
\beq
\delta_{[L/M]} \simeq
-\,\frac{M!}{K^M  }\,, \quad~{\rm where}~ K = L+M+a\,M
\label{Efive}
\eeq
and where $a$ is a number of order 1 that depends on the series under
consideration.  
For large $L,M$ eq.~\eqref{Efive} yields an 
exponential decrease of the 
the error, as in eq.~\eqref{estII}.

\begin{figure}[H]
\begin{center}
\mbox{\epsfig{file=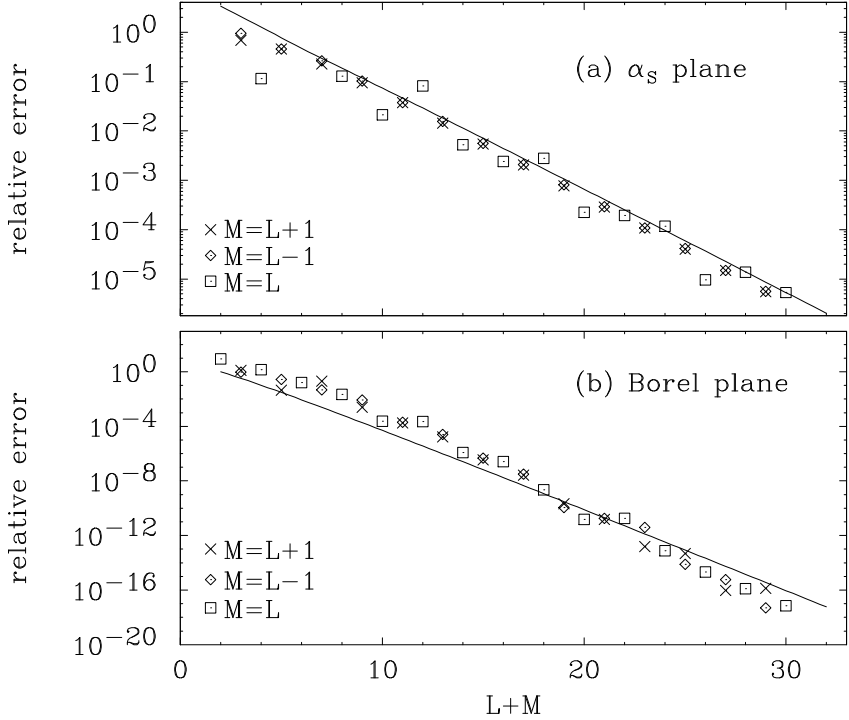,width=9.5truecm,angle=90}}
\end{center}
\caption{\small
Relative errors in the $[L/M]$ Pad\'e Approximant Predictions
\protect\cite{SEK,Erice95}
(a) for  the 
QCD vacuum polarization D-function series, evaluated to all
orders
in the large-$N_f$ approximation
\protect\cite{Dfunction} 
(the rate of convergence agrees with expectations for
a series with a discrete set of Borel poles),
and 
(b) for the Borel transform of the D-function series,
where the convergence is particularly striking.
The straight lines correspond to the exponential decay given by the
respective error formulae,
eqs.~\protect\eqref{Efive} and \protect\eqref{Eeight}.
The crosses, diamonds and squares correspond to 
$M=L+1,L+1,L$, respectively.
\label{FigIII}
}
\end{figure}

This prediction agrees very well with the known
errors in the PAP's \cite{SEK} for the QCD vacuum polarization D function
calculated in the large $N_f$ approximation \cite{Dfunction},
as seen in Fig.~\ref{FigIII}a.

One can repeat this exercise also for the Borel transform of the 
D function series. As mentioned earlier, 
a generic Borel transform is characterized by the presence of poles
(more generally, branch points) on the real axis.
In view of this, we expect an even faster convergence in this case, since
the Pad\'e, being a rational function, is particularly well-suited to 
reproduce this analytic structure.
Indeed, it turns out that in this case $\epsilon_n \sim 1/n^2$ and
\beq
\delta_{[M/M]} \simeq
-\,\frac{(M!)^2}{K^{2M}}
\label{Eeight}
\eeq
which agrees very well with the corresponding PAP results shown in 
Fig.~\ref{FigIII}b \cite{SEK,Erice95}.

The high degree of agreement between the analytical error estimates 
in eqs.~\eqref{Efive} and \eqref{Eeight} and the actual errors in PAP 
suggest that one can substantially improve the PAP method by systematically
including the error estimates $\delta_{[L/M]}$ {\em as a correction},
yielding the Asymptotic Pad\'e Approximant Predictions (APAPs):
\begin{equation}
c^{APAP}_{L+M+1} = {c^{est}_{L+M+1} \over 1 + \delta_{L+M+1}}
\label{APAP}
\end{equation}
where $c^{est}_{L+M+1}$ is the original PAP prediction without the additional
correction, as in eq.~\eqref{Efour}, and $\delta_{L+M+1}$ is obtained
by fitting \eqref{Efive} to the known lower orders
\cite{EKSbeta}.

\bigskip
The APAP method results not only in a substantial improvement of the PAP
estimates, but also significantly reduces the difference between the 
predictions based on
different $[L/M]$ values at a given order, \hbox{$L+M$=fixed}.

The method has been applied to the Bjorken sum rule for 
the difference of first moments
of proton and neutron structure functions $g_1(x,Q^2)$
in polarized deep inelastic scattering \cite{Bjcorr}. 
For $N_f = 3$ the sum rule reads
\beq
\int^1_0 [ \,g_1^p(x,Q^2) - g_1^n(x,Q^2)\,] =
{1 \over 6} |g_A|\kern-0.2em \left[ 1 - x - 3.58 x^2 - 20.22 x^3 
+  c_4 x^4 + \dots \right]
\label{bjf}
\eeq
where $x=\alpha(Q^2)/\pi$ and the exact expression for $c_4$ is still 
unknown.
\vfill\eject

The PAP  and the corresponding APAP estimates of $c_4$ are 
\beq
\begin{array}{lccc}
[2/1]:  \qquad\qquad\qquad & \tilde c_4^{PAP} \approx {-} 114 
& \quad \longrightarrow \quad & \tilde c_4^{APAP} \approx {-} 131
\\
\\
{[1/2]}: \qquad\qquad\qquad  & \tilde c_4^{PAP} \approx {-} 111 
& \quad \longrightarrow \quad & \tilde c_4^{APAP} \approx {-} 130
\end{array}
\label{c4BJSR}
\eeq
Clearly, APAP estimates show significantly less spread than the corresponding
PAP estimates.
Remarkably, the APAP estimates of $\tilde c_4$ show an
almost perfect agreement
 with an independent estimate,  based on a completely different method
\cite{KataevStarshenko}:
$\tilde c_4 = {-}130$ !

\begin{figure}[H]
\begin{center}
\mbox{\epsfig{file=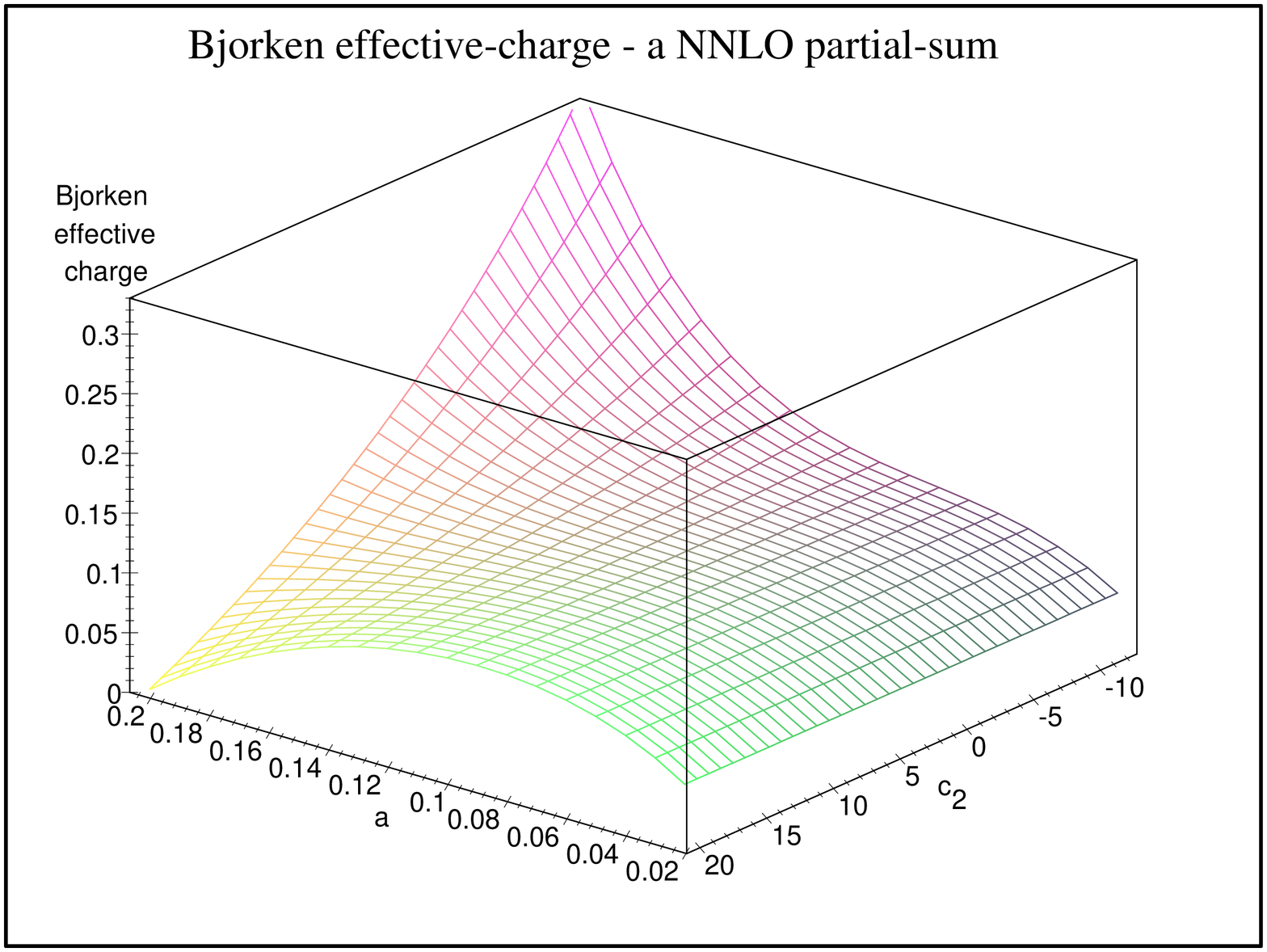,width=8.9truecm,angle=270}}
\vskip-1truecm
\mbox{\epsfig{file=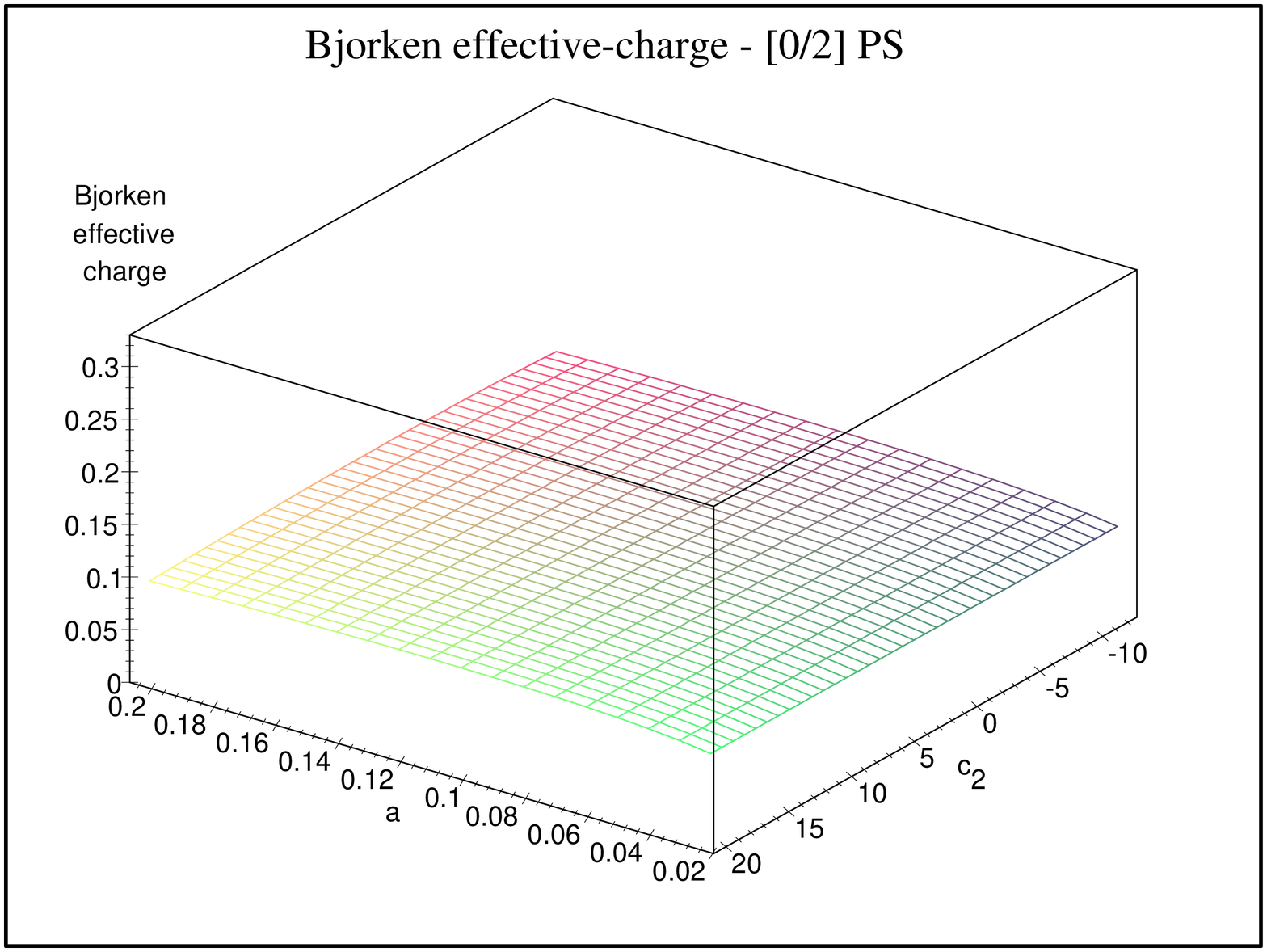,width=8.9truecm,angle=270}}
\end{center}
\caption{\small Bjorken effective charge for $Q^2\, =\, 20\,GeV^2$
plotted 
as a function of the renormalization scale and scheme, as parametrized by
the coupling \hbox{$x=\alpha_s(\mu^2)/\pi$} and the
second coefficient of the $\beta$ function: $c_2=\beta_2/\beta_0$.
The upper plot shows the NNLO partial sum, while the 
lower plot shows the [0/2] Pad\'e approximant.
\label{parcol}
}
\end{figure}

As already mentioned, a typical finite order perturbative
series such as \eqref{bjf}
exhibits a spurious renormalization scale and scheme dependence.
Schematically we have,
\beq
\underbrace{S(x)}_{
{\hbox{\footnotesize physical observable:}\atop
\hbox{\footnotesize scale and scheme}}\atop
\hbox{\footnotesize independent}
}
=
\,\,
\underbrace{c_0 + c_1 x + c_2 x^2 + c_3 x^3}_{
\hbox{\footnotesize exactly known partial sum,}\atop
\hbox{\footnotesize scale and scheme dependent}
}
\,+\,
\underbrace{c_4 x^4 + c_5 x^5 + 
\cdots}_{
\hbox{\footnotesize unknown}\phantom{q}\atop
\hbox{\footnotesize higher order terms}
}
\label{observable}
\eeq
Replacing a finite-order perturbative series by a
Pad\'e is equivalent to adding an infinite series of 
estimated terms generated by the rational approximant. 
If such an estimate is accurate, we expect to see
a reduction in the renormalization scheme and scale dependence.
As shown in Fig.~\ref{parcol},
this expectation is fully realized when Pad\'e is applied
\cite{EGKSRS} to the 
Bjorken sum rule series in eq.~\eqref{bjf}.

It turns out that this dramatic reduction in the scale and scheme
dependence can also be understood on a deeper level.
In Ref.~\cite{Einan} it was shown that in the large-$\beta_0$ limit,
i.e. when the $\beta$ function is dominated by the one loop contribution,
the scale dependence is removed completely. 
This is because
in this  limit the renormalization scale
transformation of $\alpha_s$
reduces to a homographic transformation of the Pad\'e
argument. Diagonal PA's are invariant under such transformations
\cite{Baker}.
Non-diagonal PA's are not
totally invariant, but they reduce the RS
dependence significantly \cite{Einan}.
In the real world the usual \MSbar\ $\beta$ function includes higher-order
terms beyond $\beta_0$. 
Still, in QCD with $3 \le N_f\le 5$, the 1-loop running of the coupling
is dominant and therefore  PA's are still almost invariant
under change of renormalization scale.

A further related interesting development is the observation \cite{BEGKS}
that the Pade approximant approach for resummation of perturbative
series in QCD provides a systematic method for approximating the flow
of momentum in Feynman diagrams. In the large-$\beta_0$ limit, diagonal
PA's generalize the Brodsky-Lepage-Mackenzie (BLM) scale-setting method
\cite{BLM}
to higher orders in a renormalization scale- and scheme-invariant manner.

\section{Predicting the QCD $\beta$ function at 4 and 5 loops}
Although no QCD observables have been calculated exactly beyond 
${\cal O}(\alpha^3)$, in fall of 1996 we had learned that a calculation 
of the 4-loop contribution to the QCD $\beta$ function was under way,
and likely to be published soon.
The prediction of the unknown 4-loop coefficient \cite{EKSbeta} was
therefore an important challenge and
excellent testing ground for the new APAP method.

\begin{figure}[H]
\begin{center}
\mbox{\epsfig{file=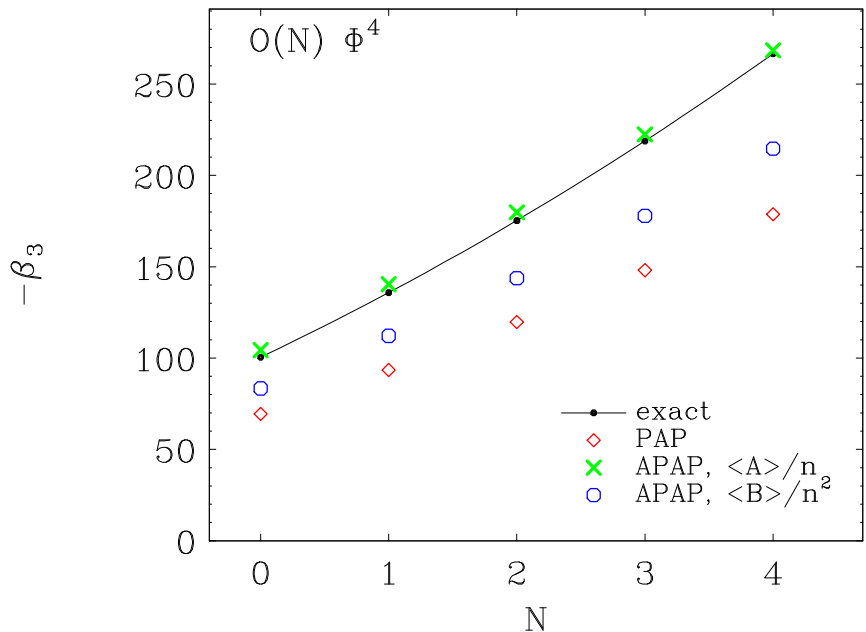,width=8.3truecm,angle=90}}
\end{center}
\caption{\small The 4-loop $\beta$-function coefficient $\beta_3$ in
$\Phi^4$ theory with $O(N)$ symmetry.
The exact results are denoted by black dots, joined by
a solid line to guide the eye. Naive PAP results are denoted by  diamonds,
and APAP results obtained from the 
$\langle A\rangle /n$ type of correction are denoted
by crosses. For comparison, also shown are
APAP results obtained from the $\langle B\rangle /n^2$ type 
of correction, denoted by open circles.
\label{betaphi4}
}
\end{figure}

As a warm-up exercise one can test the APAP method on the 4-loop 
$\beta$ function 
of the $\Phi^4$ theory with $O(N)$ global vector symmetry
\cite{vkt},
the latter being analogous to the $SU(N_f)$ global symmetry of QCD.
The results \cite{EKSbeta} for $\beta_3$ in
$O(N)$ $\Phi^4$ theory are shown in Figure \ref{betaphi4}.
Clearly, the variant of APAP method denoted as $\langle A \rangle/n$
(see \cite{EKSbeta} for details) is markedly superior to the 
naive PAP. The 5-loop $\beta$ function in this theory 
is also known \cite{Chetyrkin5loop,Kleinert5Loop} and the 
corresponding APAP estimates also turn out to be very precise
\cite{EKSbeta}.
Consequently this was the method of choice for the 
QCD 4-loop $\beta$ function.
\medskip
The strategy for computing $\beta_3$, 
the 4-loop $\beta$ function coefficient, is as follows.
We recall that $\beta_3$ is a cubic
polynomial in the number of flavors $N_f$:
\beq
\beta_3 = A_3 + B_3 N_f + C_3 N_f^2 + D_3 N_f^3,
\label{beta3form}
\eeq
where $D_3 = 1.49931$ (For $N_C=3$)
is known from large-$N_f$ calculations \cite{Gracey}.
The known exact expressions for the
1-, 2- and 3-loop $\beta$ function are used as input to APAP,
to predict the value of 
$\beta_3$ for a range of $N_f$ values.
The predictions for $A_3, B_3, C_3$ are then obtained from
fitting the APAP results for $0 \le N_f \le 4$ to a
polynomial of the form \eqref{beta3form}.

Shortly after the APAP prediction of $\beta_3$ \cite{EKSbeta}
the exact result was published in Ref.~\cite{beta4loop}. 
One important lesson from the exact results is that they
contain qualitatively new color factors,
corresponding to quartic Casimirs, analogous to light-by-light scattering
diagrams in QED. Such terms are not present at \hbox{1-,}
2- and 3-loop level, and therefore cannot be
estimated using the Pad\'e method. Numerically the correction due to these
new color factors is not very large, but in principle
the PAP estimates should be compared with the rest of the exact
expression, as is done in the first three columns of Table~I.
\hfill\break
\def\mb{\kern-0.2em}
\def\bb{\kern-1.34em}
\def\hb#1{\,\hbox{#1}\,\,}
\bigskip
\begin{table}[H]
\thicksize=0.6pt
\begintable
     | \hb{APAP}   |\hb{EXACT}|\hb{\% DIFF} |\mb|\hb{WAPAP}  |\hb{\% DIFF} \cr
$A_3$  | 23,600(900) | 24,633   | -4.20(3.70) |\mb| 24,606     | -0.11     \cr
$B_3$  | -6,400(200) | -6,375   | -0.39(3.14) |\mb|-6,374      | -0.02     \cr
$C_3$  | 350(70)     | 398.5    | -12.2(17.6) |\mb| 402.5      | -1.00      \cr
$D_3$  | \hb{input}  | 1.499    |    -        |\mb|\hb{input}  | -
\endtable
\bigskip
\caption{\small
Exact four-loop results for the QCD
$\beta$ function, compared with the original APAP's in the first column,
and improved APAP's obtained from a weighted average over negative
$N_f$ (WAPAP), as discussed in \protect\cite{EJJKS}.
The numbers in parenthesis are the error estimates
from~\protect\cite{EKSbeta}.}
\end{table}

The APAP estimates for $A_3$, $B_3$ and $C_3$ seem quite satisfactory,
until one realizes that the $A_3$ and $B_3$ terms in 
\eqref{beta3form} have opposite signs and their magnitude is such that
they almost cancel each other at $N_f\approx 4$.
This means that for numerical prediction of $\beta_3$ as function of 
$N_f$ in the physically interesting range $0 \le N_f \le 5$
a better precision is required. Fortunately, it is possible to obtain such
precision. This is accomplished by formally using negative 
values of $N_f$ in the fitting procedure, so that no cancellation occurs, 
and making a careful choice of the range of negative values of $N_f$ used 
for the fit. Once the values of $A_3$, $B_3$ and $C_3$ are obtained this
way, one can use them to compute the physical predictions 
at positive $N_f$ \cite{EJJKS}. This procedure has been referred to as 
WAPAP, for ``weighted APAP". The corresponding results are shown and
compared with exact results in the last two columns of Table I. We see 
a dramatic improvement in the precision.

\begin{figure}[H]
\begin{center}
\mbox{\epsfig{file=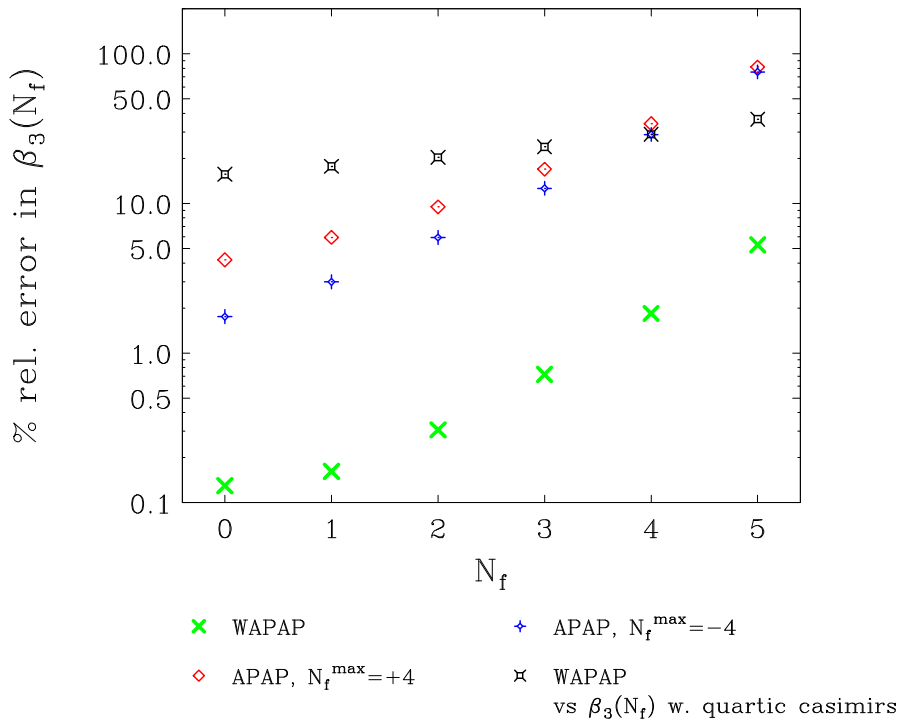,width=9.5truecm,angle=90}}
\end{center}
\caption{\small 
Predictions for $\beta_3$, as function of $N_f$,
for $N_C=3$. The percentage relative errors are obtained
using various APAP-based estimation schemes:
naive APAP's fitted with positive $N_f \le 4$
(diamonds),
naive APAP's fitted with negative \hbox{$N_f \ge -4$}, 
WAPAP's compared to the
exact value of $\beta_3$ including quartic Casimir terms, and
WAPAP's compared to $\beta_3$ without quartic Casimir
terms (crosses).
\label{beta3_error}
}
\end{figure}

Figure \ref{beta3_error} displays graphically predictions 
for $\beta_3$,
as a function of $N_f$ for the most interesting case $N_C = 3$.
We plot the percentage relative errors obtained using various APAP-based
estimation schemes \cite{EKSbeta,EJJKS}: 
naive APAP's fitted with positive $N_f \le 4$
(diamonds),
naive APAP's fitted with negative 
$N_f \ge -4$, WAPAP's compared to the
exact value of $\beta_3$ including quartic Casimir terms, and
WAPAP's compared to $\beta_3$ without quartic Casimir terms (crosses). We
see that the latter are the most accurate for $\beta_3$ in QCD.

\begin{figure}[H]
\begin{center}
\mbox{\epsfig{file=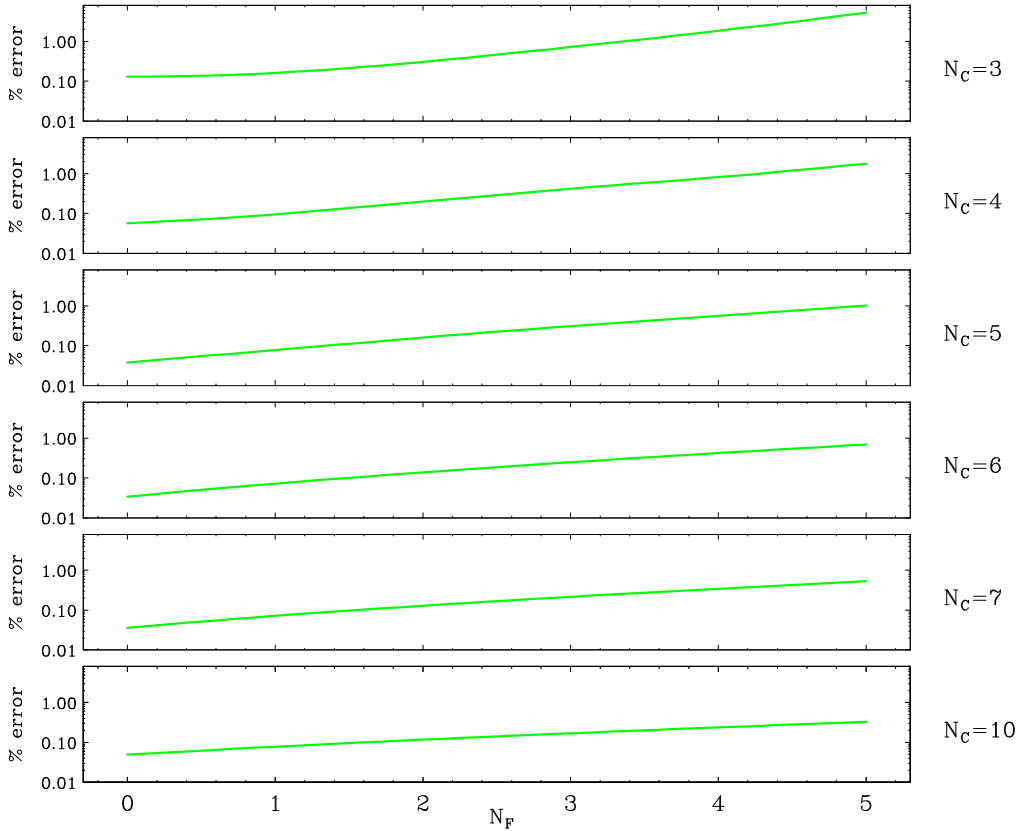,width=11.0truecm,angle=90}}
\end{center}
\caption{\small The percentage relative errors in the WAPAP prediction
for $\beta_3$ (compared to the exact result with quartic Casimir terms
omitted), plotted vs. $N_f$
for $N_C=$3, 4, 5, 6, 7, 10.
\label{beta3_vs_Nf_vs_Nc}
}
\end{figure}

In Figure \ref{beta3_vs_Nf_vs_Nc} we show the error in the  
WAPAP prediction for $\beta_3$
as a function of $N_f$, for $N_C=$3, 4, 5, 6, 7 and 10,
once again omitting quartic Casimir terms from the exact result.
The accuracy of these predictions is our best evidence for
believing in the utility of the WAPAP method.

%
%
%
%

The WAPAP method does very well on the four-loop QCD $\beta$ function, 
but the details of the method were fine-tuned after the exact results 
became available. In Ref.~\cite{EJJKS} predictions were also given for
yet unknown 
5-loop $\beta$ function in QCD and 4- and 5-loop 
$\beta$ function in $N=1$ supersymmetric QCD.
It is extremely interesting to see how well
our predictions will do for these quantities.
\vfill\eject

\noindent
{\bf Acknowledgments}

This paper is devoted to the memory of Mark Samuel, a wonderful collaborator
and friend, who passed away suddenly last November. The results described
here were obtained in a close collaboration with him, together with
Stan Brodsky, John Ellis, Einan Gardi, Ian Jack and Tim Jones.

This research 
was supported in part by the Israel Science Foundation
administered by the Israel Academy of Sciences and Humanities,
and by
a Grant from the G.I.F., the
German-Israeli Foundation for Scientific Research and
Development.
\def\PL{{\sl Phys. Lett.\ }}
\def\NP{{\sl Nucl. Phys.\ }}
\def\PR{{\sl Phys. Rev.\ }}
\def\PRL{{\sl Phys. Rev. Lett.\ }}

\end{document}